\documentclass{article}

\usepackage{arxiv}

\usepackage[utf8]{inputenc} 
\usepackage[T1]{fontenc}    
\usepackage{hyperref}       
\usepackage{url}            
\usepackage{booktabs}       
\usepackage{amsfonts}       
\usepackage{nicefrac}       
\usepackage{microtype}      
\usepackage{lipsum}		
\usepackage{graphicx}
\usepackage{natbib}
\usepackage{doi}

\title{Decentralized Deepfake Detection Blockchain Network using Dynamic Algorithm management}

\author{ \href{https://orcid.org/0000-0001-5431-6367}{\includegraphics[scale=0.06]{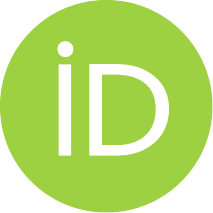}\hspace{1mm}Dipankar Sarkar} \\
  Cryptuon Research \\
  \texttt{me@dipankar.name} \\
}

\renewcommand{\headeright}{}

\hypersetup{
pdftitle={Decentralized Deepfake Detection Blockchain Network using Dynamic Algorithm management},
pdfsubject={cs.CR},
pdfauthor={Dipankar Sarkar},
pdfkeywords={deepfake, deep learning, blockchain, decentralisation, trustless, media, algorithms, dynamic},
}

\begin{document}
\maketitle

\begin{abstract}
Deepfake technology is a major threat to the integrity of digital media. This paper presents a comprehensive framework for a blockchain-based decentralized system designed to tackle the escalating challenge of digital content integrity. The proposed system integrates advanced deep learning algorithms with the immutable and transparent nature of blockchain technology to create a trustless environment where authenticity can be verified without relying on a single centralized authority. Furthermore, the system utilizes smart contracts for dynamic algorithm management and token-based incentives further enhances the system's effectiveness and adaptability. The decentralized architecture of the system democratizes the process of verifying digital content and introduces a novel approach to combat deepfakes. The collaborative and adjustable nature of this system sets a new benchmark for digital media integrity, offering a more robust digital media environment.
\end{abstract}


\section{Introduction}
In the digital age, the accuracy of digital media has become a key factor in reliable communication. The emergence of deepfake technology, which uses advanced artificial intelligence (AI) to alter audio and video material, presents an unprecedented challenge to the integrity of information. Deepfakes can undermine public discourse, endanger privacy rights, and disrupt democratic processes by creating convincingly realistic media content.

The rapid development of deepfakes has made the problem worse. As detection methods become more advanced, so do the techniques used to create more complex forgeries, resulting in a competition between forgers and detectors. Traditional centralized methods of content verification are unable to keep up with the vast amount of content and the complexity of these manipulations. Therefore, there is an urgent need for a decentralized, scalable, and constantly evolving system that can verify digital media.

This paper presents a new blockchain-based decentralized system that is intended to tackle the issue of deepfake detection. Our system takes advantage of the immutable and transparent features of blockchain technology to create a trustless environment where the authenticity of data can be confirmed without relying on a single centralized entity. Additionally, the system uses smart contracts to motivate the development and implementation of the most advanced deepfake detection algorithms. Our contribution is threefold:

\begin{enumerate}
    \item \textbf{Decentralization} We propose a decentralized architecture that \textit{democratizes the verification process}, allowing multiple stakeholders to participate in maintaining the integrity of digital media.
    \item \textbf{Adaptability} Recognizing the dynamic nature of AI-generated content manipulation, our system is designed to integrate \textit{new detection algorithms} as they emerge. Smart contracts automatically manage the life cycle of these algorithms within the network, ensuring that only the most effective models are utilized.
    \item \textbf{Incentivization} To encourage the continuous improvement of deepfake detection technologies, our system rewards contributors with tokens based on the performance of their algorithms. This not only fosters a competitive environment for innovation, but also ensures that the system remains at the\textit{forefront of detection capability}.
\end{enumerate}

The rest of this paper is structured as follows: Section 2 reviews the existing literature on deepfake detection and its shortcomings. Section 3 outlines the theoretical basis of our proposed system. Section 4 outlines the system architecture, and Section 5 covers the deployment of the algorithms. Finally, we discuss the implications of our work and potential directions for future research.

\section{Related Work}

The digital content verification landscape is constantly changing as artificial intelligence (AI) progresses. To detect deepfakes, a variety of techniques have been used, from traditional digital watermarking to advanced machine learning (ML) algorithms. Research in this area suggests that blockchain's decentralized trust mechanisms and AI's analytical capabilities should be combined for deepfake detection. However, this integration requires new methods of algorithm management and incentive models to ensure a high level of adaptability and responsiveness to the ever-evolving deepfake technology.

\subsection{Deepfake Detection}

The initial strategies for detecting manipulated content were heavily based on manual verification by experts. These techniques often included analysis of visual artifacts or inconsistencies within the content. \citep{9721302} provided comprehensive literature review on deepfake detection summarizing various methodologies, a systematic review covered relevant articles from 2018 to 2020.

With the rise of deepfakes, researchers quickly turned to machine learning for automated detection. \citep{masood2021deepfakes} has detailed analysis of existing tools and ML approaches for deepfake generation and detection, presented in a state-of-the-art review, focusing on both audio and visual deepfakes.

\citep{rössler2019faceforensics} made significant strides with the introduction of the FaceForensics++ benchmark, which provided a data set to train and evaluate deepfake detection models. This data set became a foundational resource for future research in the field.

\subsection{Blockchain - Trust Mechanism}

In parallel to advances in deepfake detection, blockchain technology has been explored as a means of establishing trust in digital transactions. The immutable nature of the blockchain makes it a powerful tool for verifying the authenticity of digital assets, as discussed by \citep{GUO2022100067} in a survey on blockchain technology and its security, which compared various consensus algorithms and explored blockchain applications.

The fusion of blockchain and AI for content verification is a relatively new area of research. A systematic review of blockchain research done by \citep{doi:10.1080/17517575.2021.1939895} during the period 2016 to 2020 highlighted the technological characteristics and blockchain implementation models, setting the stage for further exploration in the context of deepfake detection.

The use of smart contracts to dynamically manage and incentivize the performance of deepfake detection algorithms is an innovative approach that has not been extensively explored in existing literature. However, the concept of using smart contracts to facilitate decentralized autonomous organizations (DAOs) in other domains has been discussed in \citep{10.1371/journal.pone.0258995} investigating how blockchain-based applications might affect firms' organizations and strategies by 2030.

\subsection{Challenges and Limitations}

Despite these developments, challenges remain, particularly in the scalability and real-time application of detection methods. Current detection algorithms struggle to keep pace with the rapid improvements in deepfake generation, a problem highlighted in a comprehensive study by \citep{wang2023deepfake} from the reliability perspective of deepfake detection.

Furthermore, the ability to detect re-recorded or subtly altered content has been a persistent challenge. The re-recording attack vector is particularly problematic, as it introduces anomalies that automated systems can overlook, as discussed in an analysis of deepfake detection techniques \citep{Nguyen_2022}, which examined the use of sophisticated machine learning algorithms to create falsified media content. Meanwhile, accommodating legitimate alterations to content, such as cropping or resizing, without triggering false detections, is a delicate balance that requires further research.

\section{Theoretical Framework}

This paper examines the theoretical basis of the proposed system, which integrates blockchain technology with AI for deepfake detection. The framework combines elements from computer science, cryptography, and machine learning to create a secure, decentralized solution for verifying digital media. The combination of blockchain and AI, with the help of smart contract technology, offers a promising way to address the difficulties posed by deepfakes. This system seeks to create a decentralized, transparent, and adaptive approach to authenticating digital content by utilizing these technologies.

\paragraph{Blockchain Technology} Blockchain serves as the backbone of the proposed system, providing a decentralized and immutable ledger to record transactions and data. The work of \citep{nakamoto2009bitcoin} on Bitcoin introduced the concept of a decentralized ledger, laying the foundation for subsequent blockchain applications. Recent advances demonstrate the potential for blockchain to ensure data integrity and transparency in various domains, including verification of digital content.

\paragraph{Smart Contracts} Smart contracts are self-executing contracts with the terms of the agreement directly written in code. They are crucial for automating the management of deepfake detection algorithms within our system. \citep{Wu2022} discussed the evolution of smart contract technology and its increasing role in facilitating complex decentralized applications.

\paragraph{Deep Learning for Deepfake Detection} Deep learning has become a powerful tool for detecting deep-fakes. \citep{passos2023review} provide a comprehensive review of deep learning techniques used in biometric security, including deepfake detection. The adaptability of these models is crucial to keep up with the evolving nature of deep-fake technology.

\paragraph{Integrating AI with Blockchain} The integration of AI and blockchain for content verification is a novel approach that has been gaining traction. A study by \citep{Wang2021TheAO} highlighted the potential of this integration to improve data security and trust in AI applications. They further discuss how blockchain can provide a framework for the reliable deployment of AI models, including those used in deep-fake detection.

\paragraph{Algorithm Management and Dynamic Updating} The dynamic management of algorithms is a critical aspect of the proposed system. The work of \citep{vyas2022} on the use of blockchains to manage AI agents in EHR systems emphasizes the importance of a flexible and adaptable system in response to rapidly evolving AI technologies.

\section{System Architecture}

\begin{figure}
	\centering
        \includegraphics[width=15cm]{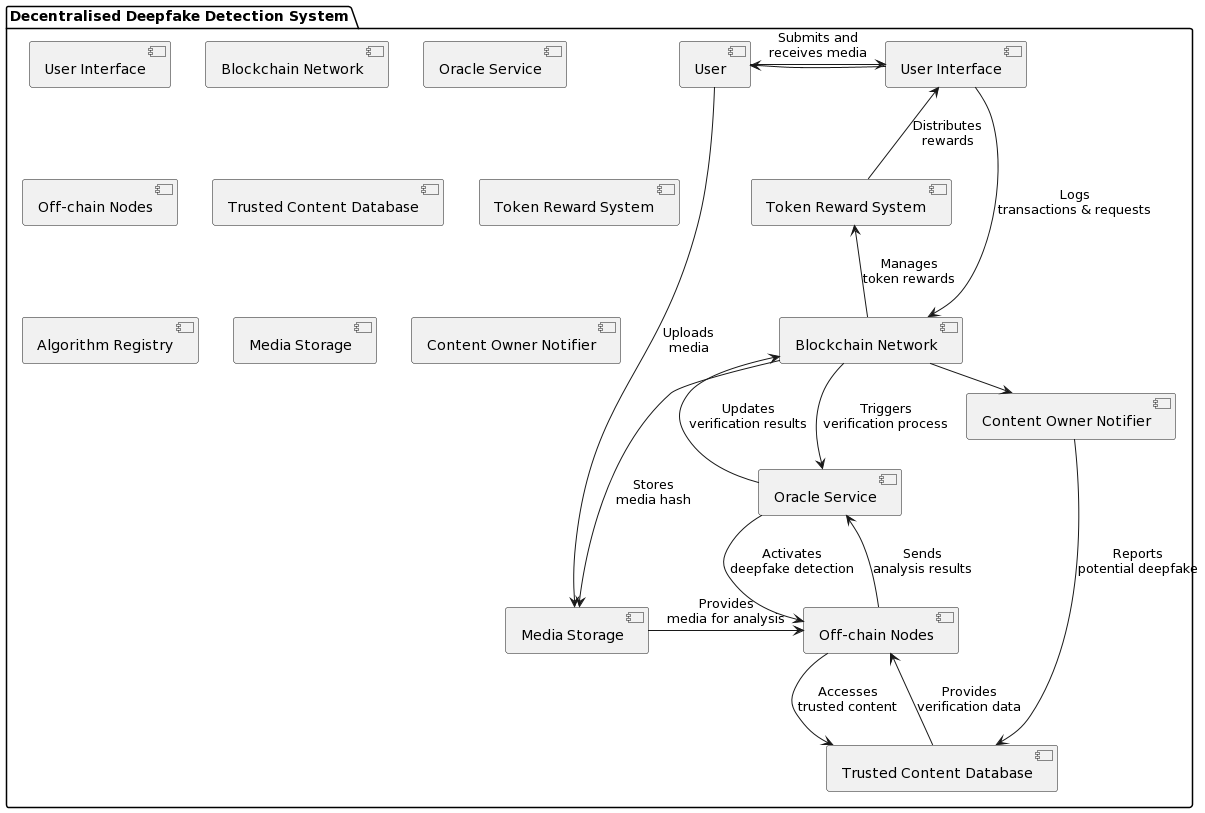}
        \caption{System flow}
	\label{fig:fig1}
\end{figure}

This system architecture incorporates blockchain technology, smart contracts, deep learning models, and a trusted content provider framework to create a comprehensive, decentralized solution for deepfake detection. This multi-faceted strategy guarantees the system's efficacy, flexibility, and scalability.

This architecture offers a comprehensive and dynamic approach to deepfake detection, combining the transparency and immutability of blockchain technology with sophisticated AI techniques and a reliable content provider framework. The system is designed to be adjustable, scalable, and efficient, addressing the difficulties posed by the ever-changing landscape of digital content manipulation.

\subsection{Blockchain Network}

The blockchain network serves as a decentralized ledger that records all transactions and data associated with deepfake detection. These are the general configurations.

\begin{itemize}
    \item \textbf{Nodes}  Nodes are categorized into full nodes, which maintain a complete copy of the blockchain, and specialized nodes, such as those responsible for validating transactions or hosting deep learning models.
    \item \textbf{Consensus Mechanism} A Proof of Stake (PoS) model is employed for consensus, balancing energy efficiency and security.
    \item \textbf{Data Storage} The blockchain stores cryptographic hashes of digital content, results from deepfake analysis, and references to trusted content provided by registered entities.
\end{itemize}

\subsection{Smart Contracts}

Smart contracts are utilized to automate essential operations within the network. These are the primary functions that are implemented using them.

\begin{itemize}
    \item \textbf{Algorithm Management} These contracts handle the registration, validation, and performance tracking of deepfake detection algorithms.
    \item \textbf{Trusted Content Registry} Smart contracts also manage a trusted content registry, including metadata and hashes, provided by content creators or entities.
    \item \textbf{Incentive Distribution} Token rewards are distributed based on algorithm performance and contributions from nodes or content providers.
\end{itemize}

\subsection{Deep Learning Models}

Deep learning models are central to detecting deepfakes:

\begin{itemize}
    \item \textbf{Model Hosting and Execution} Models are hosted and executed off-chain to enhance computational efficiency, with results communicated back to the blockchain.
    \item \textbf{Integration with Trusted Content} Some models leverage the trusted content dataset for improved accuracy, especially in detecting sophisticated deepfakes.
\end{itemize}

\subsection{Trusted Content Provider Framework}

A vital component of the system is the trusted content provider framework:

\begin{itemize}
    \item \textbf{Content Registration} Content providers can submit original content, which is hashed and stored in the blockchain. This creates a reference dataset for deepfake detection and notification.
    \item \textbf{Notification Mechanism} Providers are notified if their content is flagged as part of a deepfake, enhancing their ability to manage their digital reputation.
\end{itemize}

\subsection{User Interface}

The system includes a user interface for interaction with various stakeholders:

\begin{itemize}
    \item \textbf{Content Submission and Retrieval} Users submit content for analysis and access results, while content providers manage their trusted content.
    \item \textbf{Interfacing with Blockchain} The interface communicates with the blockchain for submitting and retrieving data, maintaining user-friendly accessibility.
\end{itemize}

\subsection{Oracle Service}

The oracle service links the off-chain and on-chain components:

\begin{itemize}
    \item \textbf{Analysis Activation} The service activates the deep learning models for content analysis upon request.
    \item \textbf{Result Communication} After analysis, the oracle communicates the results to the blockchain for record-keeping and user access.
\end{itemize}

\subsection{Tokenomics}

The token economy incentivizes participation and sustains the ecosystem:

\begin{itemize}
    \item \textbf{Rewards System} Contributors, including algorithm developers, nodes, and content providers, receive tokens for their contributions.
    \item \textbf{Transaction Facilitation} Tokens are used for various transactions, such as accessing advanced detection features.
\end{itemize}

\section{Algorithm Deployment}

\begin{figure}
	\centering
        \includegraphics[width=10cm]{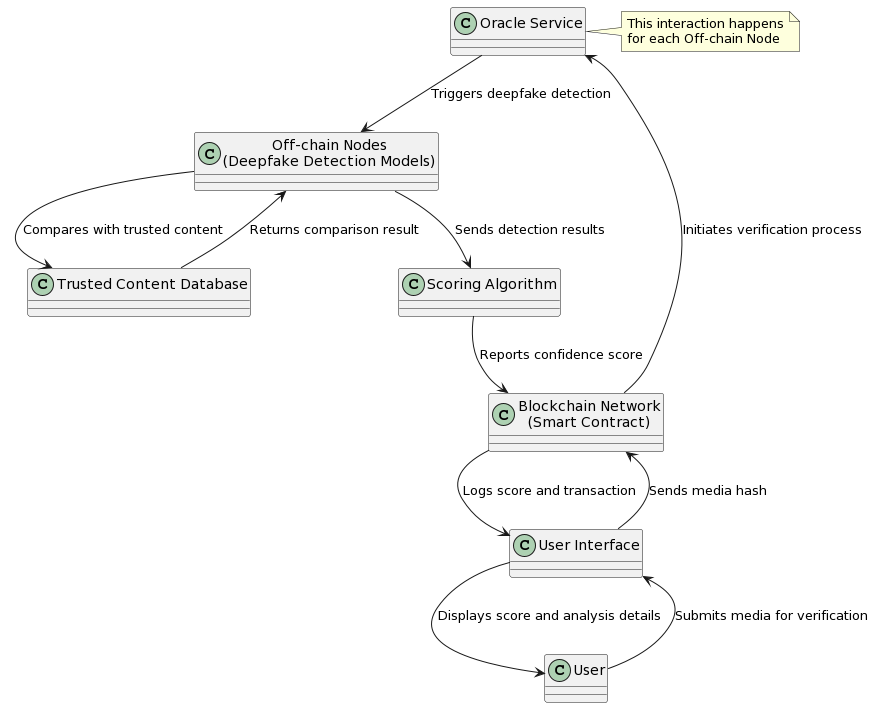}
        \caption{Process for media analysis}
	\label{fig:fig2}
\end{figure}

The quick advancement of multimedia material and the abundance of deepfakes necessitates a tactical and adjustable strategy for the implementation of the most up-to-date deepfake detection algorithms in a blockchain-based system. This section explains how audio, images, and video are managed with the help of sophisticated models, and provides links to the most recent studies in the area.

The proposed system is designed to deploy algorithms dynamically, efficiently, and responsively to the changing nature of deepfakes. By taking advantage of the most recent developments in deep learning and AI, as well as a strategic approach to managing different media types, the system is intended to offer a strong defense against the increasing danger of deepfake technology.

\subsection{Different Media Types}
\subsubsection{Images and Video}
\paragraph{Detection Models} Utilize convolutional neural networks (CNNs) and generative adversarial networks (GANs) for image and video analysis, as these models excel in processing visual data. As suggested by \citep{tolosana2020deepfakes} these models can effectively identify various deepfake techniques, including face modification and expression swaps.

\paragraph{Embedding Generation} For searching and comparison, models like those discussed by \citep{afchar2018mesonet} and \citep{li2021multi} can be used to generate embeddings that represent the unique features of each piece of content, helping to identify deep-fakes.

\subsubsection{Audio}
\paragraph{Analysis Models} Deploy models that specialize in audio processing, such as recurrent neural networks (RNNs) and their variants, which are adept at handling sequential data. This approach is informed by the work of \citep{deng2019exploiting}, who exploited time-frequency patterns with LSTM-RNNs for low-bitrate audio restoration.

\paragraph{Comparison with Trusted Datasets} Use pre-trained models for voice recognition and comparison against trusted datasets, to detect impersonation and audio manipulation. This approach aligns with the work of \citep{radford2021robust}, who demonstrated the robustness of speech recognition via large-scale weak supervision.

\subsection{Model Deployment Strategy}

\paragraph{Model Selection and Updating} The system dynamically selects the most suitable model based on the type of media and the specific characteristics of the content. This is informed by the research of \citep{masood2021deepfakes}, which highlights the importance of customized strategies for different fake techniques. Models are continuously updated and refined through a feedback loop, using performance metrics and user feedback to inform improvements.

\paragraph{Off-Chain Execution for Efficiency} Following the analysis by \citep{sariboz2021offchain}, off-chain execution is used to efficiently manage computational demands, particularly when dealing with large data sets and complex models.

\paragraph{Incorporation of Transfer Learning} Utilize transfer learning to accelerate model deployment, as it allows to leverage pre-trained models to quickly adapt to new deepfake techniques. This method is in line with the insights provided by \citep{ijgi10030137}.

\subsection{Embedding and Search Mechanism}

\paragraph{Generating Embeddings} Implement models to generate distinct embeddings for each piece of content, allowing quick search and comparison against the database of trusted content and previously analyzed media.

\paragraph{Cross-Referencing with Trusted Content} When a piece of content is submitted, its embeddings are compared with those of the trusted content database to identify potential matches or manipulations.

\subsection{Addressing the Challenge of Evolving Deepfakes}

\paragraph{Adaptive Models} The system incorporates the latest developments in AI and machine learning to adapt to new deep-fake techniques, as the technology continues to advance, as described by \citep{castillo2021comprehensive}.

\paragraph{Community Contributions and Model Validation} Encourage community contributions of new models, which are validated using the blockchain network and smart contracts to ensure reliability and effectiveness.

\section{Discussion}

This section offers an in-depth examination of the proposed blockchain-based deepfake detection system, exploring its advantages, drawbacks, potential improvements, and ethical implications, as well as addressing any implementation difficulties.

This proposed blockchain-based deepfake detection system is a major advancement in tackling the difficulties posed by deepfake technology. It is designed to stay up-to-date and relevant by continually adapting and incorporating the latest advancements in blockchain and AI. To ensure the responsible and effective use of this technology, ethical considerations and a dedication to resolving implementation issues are essential.

\subsection{System Strengths}

\begin{figure}
	\centering
        \includegraphics[width=15cm]{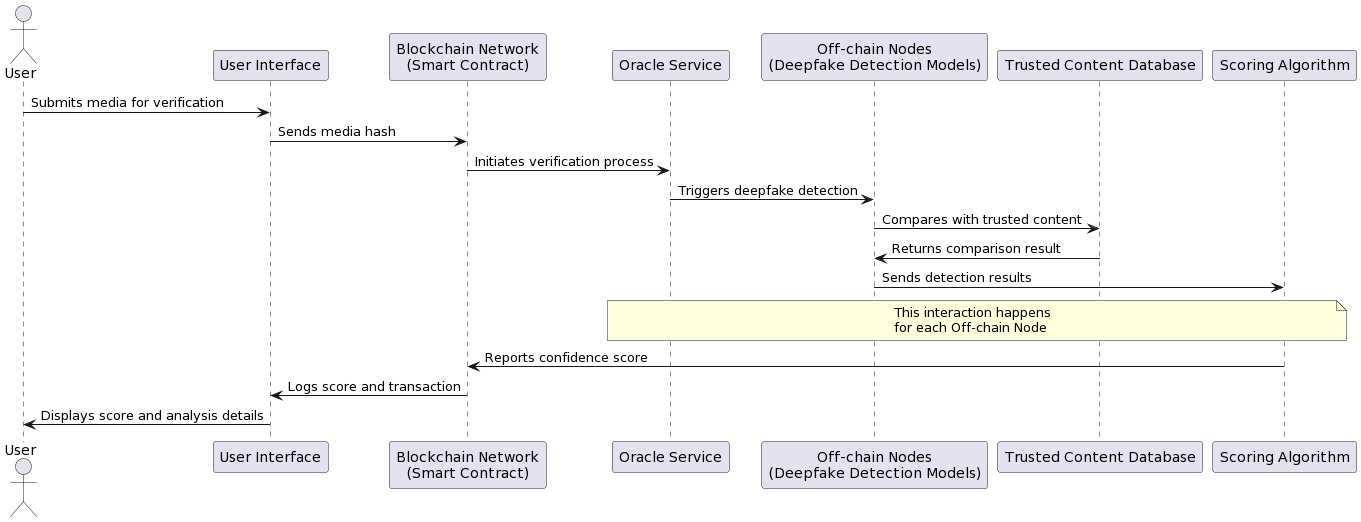}
        \caption{System process diagram}
	\label{fig:fig3}
\end{figure}

The main advantage of the proposed system is its creative combination of blockchain technology and sophisticated deep learning algorithms for the identification of deepfakes. This integration offers several key benefits:

\begin{enumerate}
    \item \textbf{Decentralization and Transparency} The blockchain framework ensures a decentralized and transparent process to verify digital content, mitigating risks associated with centralized control and potential biases.
    \item \textbf{Dynamic Adaptability} The system's ability to dynamically update and integrate new detection algorithms keeps it agile and responsive to evolving deepfake technologies.
    \item \textbf{Incentivization and Participation} The token-based incentive model encourages continuous participation and contribution from a broad community, fostering a collaborative environment for deepfake detection.
    \item \textbf{Privacy and Security} The use of cryptographic hashes and secure data storage mechanisms protects the privacy and integrity of the content being analyzed.
\end{enumerate}

\subsection{Limitations and Future Work}

This system offers a promising solution, however, it has certain restrictions that point to areas for further research.

\begin{itemize}
    \item \textbf{Scalability Concerns} As the volume of content and the number of participating nodes increase, scalability could become a challenge. Future work will involve optimizing the network architecture and improving the efficiency of algorithms to handle larger scales.
    \item \textbf{Real-Time Processing} The current framework may not support real-time analysis of media content, which is crucial for timely detection. Future enhancements will focus on reducing latency and improving processing speeds.
    \item \textbf{Diverse Data Representation} Ensure that the system effectively handles diverse forms of media across different cultures and languages is an ongoing challenge that requires continuous enhancement of the underlying models and algorithms.
    \item \textbf{Re-recording Attack Vector} Re-recording attacks, where a deepfake is recorded from a screen, can introduce anomalies that standard detection algorithms might miss. Implementing specialized algorithms that detect these specific anomalies and integrate them into the system is crucial to address this challenge.
    \item \textbf{Handling Legitimate Content Alterations} Distinguishing between maliciously altered content and legitimately modified media (e.g., cropping, resizing) is complex. Developing more sophisticated models capable of understanding the context and intent behind the alterations is key. This involves training algorithms on diverse datasets that include examples of both legitimate modifications and malicious tampering.
\end{itemize}

\subsection{Ethical Considerations}

The ethical implications of deepfake detection technology are of utmost importance. It is essential to consider the potential implications of technology on privacy, accuracy, and fairness when developing and deploying fake detection systems. Additionally, it is important to consider the potential for misuse of the technology and the potential for unintended consequences of its use.

\begin{itemize}
    \item \textbf{Bias and Fairness} It is essential to guarantee that algorithms do not display any prejudice against certain groups or individuals, which requires frequent reviews for impartiality and prejudice.
    \item \textbf{Misuse of Technology} The technology that is designed to detect deepfakes carries the potential for it to be utilized to generate more advanced deepfakes. To reduce this danger, it is essential to put in place stringent access regulations and ethical standards.
    \item \textbf{Impact on Privacy and Consent} The system must be able to tackle the intricate matters related to privacy and permission, especially when examining material that includes people who have not given their approval for such examination.

\end{itemize}

\section{Conclusion}

This paper proposed a decentralized system based on blockchain technology to tackle the growing issue of deepfake detection. By combining sophisticated deep learning algorithms with the immutable and transparent features of blockchain, a novel approach to authenticating digital media was presented. Smart contracts for dynamic algorithm management and token-based rewards further improved the system's efficiency and flexibility.

The decentralized architecture of the system democratizes the process of verifying digital content and introduces a novel approach to combat deepfake technology. The system's adaptability, which allows for the integration and updating of detection algorithms, ensures it is able to keep up with the ever-changing landscape of digital manipulation. Additionally, the token-based incentive model encourages collaboration and participation, stimulating ongoing improvement and community involvement.

Despite its advantages, the system is confronted with certain restrictions and difficulties, such as scalability problems, the requirement for instantaneous processing capacities, and the management of varied data. Moreover, the system must traverse the intricate ethical terrain that encompasses deepfake detection, including worries about prejudice, confidentiality, and the potential misuse of technology. Addressing these restrictions and ethical considerations will be essential for the successful implementation and functioning of the system.

Future research will concentrate on refining the system's design to increase scalability and effectiveness, creating algorithms that can do real-time analysis, and making sure the system can adjust to a variety of media types and situations. It will also be essential to have ethical rules and regular reviews for prejudice and equity to guarantee that the technology is used properly.

To sum up, the blockchain-based system for deepfake detection proposed in this paper is a major step forward in preserving the reliability of digital media. As deepfakes remain a danger to the trustworthiness of information, systems such as this one are essential for protecting the genuineness of digital material. The collaborative and adjustable nature of this system establishes a new benchmark for deepfake detection, offering a more robust digital media environment.

\bibliographystyle{unsrtnat}
\bibliography{references} 
\end{document}